\documentclass[10pt,twocolumn,letterpaper]{article}
\pdfoutput=1
\usepackage{cvpr}
\usepackage{times}
\usepackage{epsfig}
\usepackage{graphicx}
\usepackage{url}
\usepackage{amsmath}
\usepackage{amssymb}
\usepackage{multirow}
\usepackage{booktabs}       % professional-quality tables

\usepackage[pagebackref=true,breaklinks=true,letterpaper=true,colorlinks,bookmarks=false]{hyperref}

\cvprfinalcopy % *** Uncomment this line for the final submission

\def\cvprPaperID{****} % *** Enter the CVPR Paper ID here

\ifcvprfinal\fi
\begin{document}

%%%%%%%%% TITLE
\title{DaST: Data-free Substitute Training for Adversarial Attacks}

\author{Mingyi Zhou$^{1,2,}$\footnotemark[1]  , Jing Wu$^{1,2,}$\thanks{Equal contribution}  , Yipeng Liu$^{1,}$\thanks{Corresponding author}  , Shuaicheng Liu$^{1,2}$, Ce Zhu$^{1}$\\
%School of Information and Communication Engineering,\\
University of Electronic Science and Technology of China$^1$\\
Megvii Technology$^2$\\
{\tt\small \{zhoumingyi,wujing\}@std.uestc.edu.cn,\{yipengliu,liushuaicheng,eczhu\}@uestc.edu.cn}
%\and
}

\maketitle

\begin{abstract}
	Machine learning models are vulnerable to adversarial examples. For the black-box setting, current substitute attacks need pre-trained models to generate adversarial examples. However, pre-trained models are hard to obtain in real-world tasks. In this paper, we propose a data-free substitute training method (DaST) to obtain substitute models for adversarial black-box attacks without the requirement of any real data. To achieve this, DaST utilizes specially designed generative adversarial networks (GANs) to train the substitute models. In particular, we design a multi-branch architecture and label-control loss for the generative model to deal with the uneven distribution of synthetic samples. The substitute model is then trained by the synthetic samples generated by the generative model, which are labeled by the attacked model subsequently. The experiments demonstrate the substitute models produced by DaST can achieve competitive performance compared with the baseline models which are trained by the same train set with attacked models. Additionally, to evaluate the practicability of the proposed method on the real-world task, we attack an online machine learning model on the Microsoft Azure platform. The remote model misclassifies 98.35\% of the adversarial examples crafted by our method. To the best of our knowledge, we are the first to train a substitute model for adversarial attacks without any real data. Our codes are publicly available \footnote{{\url{https://github.com/zhoumingyi/DaST}}}. 
\end{abstract}

%%%%%%%%% BODY TEXT
\section{Introduction}
Deep neural networks have been shown vulnerable to examples with imperceptible perturbations~\cite{szegedy2013intriguing}. This causes researchers a high interest in studying attacks and defenses for assessing and improving the robustness of networks. Adversarial attack methods can be categorized into two main attacks, white-box attacks that have full access to the attacked model and black-box attacks that have partial information of models.

Black-box attacks are more practical in real world systems compared with white-box attacks. Among these attacks, score-based attacks~\cite{chen2017zoo,ilyas2018black,ilyas2018prior,guo2019simple} and decision-based attacks~\cite{brendel2017decision,cheng2018query,chen2019hopskipjumpattack} directly attack the attacked model using class probabilities or hard labels returned by the attacked model. These attack methods do not need a pre-trained substitute model, however, as a cost, they need numerous queries for attacked models to generate each attack.

Instead, gradient-based attack methods~\cite{goodfellow6572explaining,kurakin2016adversarial,papernot2016limitations,moosavi2016deepfool} need knowledge of the architecture and weights of the attacked models. Goodfellow \etal~\cite{goodfellow6572explaining} showed that adversarial examples have the property of transferability which means that adversarial examples generated for one model through white-box attack methods can also attack other models. Therefore, to carry out attack methods in the black-box setting, they use a substitute model to find the adversarial examples and then attack the machine learning model based on the transferability of these adversarial examples. Compared with current score-based and decision-based attacks, the substitute attacks do not need queries for generating adversarial examples. However, they need a pre-trained model to generate adversarial attacks. Papernot \etal~\cite{papernot2017practical} developed a method that uses a number of images to imitate the outputs of attacked models to obtain substitute networks. Prediction APIs were also developed to steal the machine models~\cite{tramer2016stealing}. Orekondy \etal~\cite{orekondy19cvpr} proposed a "knockoff" to steal the function of machine learning models. These methods do not need a pre-trained model but require many real data labeled by the attacked model to train a substitute model. However, the real images are hard to get in some real-world tasks. Therefore, it is important to develop a data-free substitute attack, such that the risks faced by current machine learning models can be assessed more comprehensively.  

In this study, we propose a data-free substitute training (DaST) method to train a substitute model for adversarial attacks. We utilize generative adversarial networks (GANs) to create synthetic samples to train the substitute model. The substitute model uses these samples to train, where the labels of the samples are produced by the attacked model. For the performance, the synthetic samples should be equally distributed in the input space. The label of samples should span all categories. However, conventional GANs without real data may generate samples that have extremely uneven distribution and only contain few categories, which means the substitute model cannot learn the classification characteristics of the attacked model comprehensively. 

To address this problem, we design a multi-branch architecture and a label-control loss for the generative model to deal with the uneven distribution of synthetic samples. The generative model can produce synthetic samples with random labels given by the attacked model. As such, the substitute model can learn the classification characteristics of the attacked model in this adversarial training and produce adversarial examples which have strong transferability for the attacked model. The main contributions of this study are summarized as follows:
\begin{itemize}
	\item We are the first to train a substitute model for adversarial attacks without any real data. Attackers can use this method to train a substitute model for adversarial attacks without collecting any real data.
	
	\item We evaluate the effectiveness of DaST on both local deep learning models and the online machine learning system, which reveals a fact that the current machine learning model has significant risks to be attacked.
	
	\item we evaluate the performance of our method in two attack scenarios, including a probability-only scenario that attacker can access the output probability of the attacked model, and a label-only scenario that attacker only accesses the output label of the attacked model. Our method generates adversarial examples efficiently on both scenarios.
\end{itemize}
In addition, we use different model architectures for the substitute model to test the influence on attack success rate caused by the model capacity.

The rest of our paper is organized as follows: in section 2, we introduce the related works. The proposed method is described in section 3.  We evaluate the performance of DaST in section 4. 

\section{Related Works}

\paragraph{Adversarial Scenes}
Adversarial attacks are carried out in the white-box setting or the black-box setting. In the white-box setting, the attacker has access to the structure and weights of the attacked models. On the contrary, in the black-box setting, the attacker only has the substitute model (gradient-based attacks) or access the outputs returned by the attacked models (query-based attacks). Black-box attack methods are more practical on real tasks.
\vspace{-1em}
\paragraph{Adversarial Attacks}
Gradient-based attacks such as FGSM~\cite{goodfellow6572explaining} and BIM~\cite{kurakin2016adversarial} have full access to the models, so they usually use a pre-trained substitute model to generate adversarial examples, and then attack the attacked model using the transferability of adversarial examples. FGSM aims to find adversarial examples by directly increasing the loss of the model, BIM is an iterative version of FGSM. Likewise, DeepFool~\cite{moosavi2016deepfool} finds adversarial examples that are likely to cross the decision boundary. To find perturbations with minimal $\ell_p$ norm, Nicholas Carlini and David Wagner~\cite{carlini2017towards} introduced a method to craft these perturbations through simultaneously minimizing the perturbations. Similar to this method, Rony \etal~\cite{rony2018decoupling} also constrain the  $\ell_2$ norm of the perturbations, they decoupled the value and direction of the perturbation. In the black-box setting, these attacks rely on the transferability of adversarial examples. However, Liu \etal~\cite{liu2017delving} showed that these examples nearly have no transferability on attacked attacks. Instead, Cheng \etal~\cite{chen2017zoo} proposed a score-based attack method zeroth order based attack (ZOO) using gradient estimation, and Ilyas \etal~\cite{ilyas2018prior} improve the way of the gradient estimation. Instead of gradient estimation, Guo \etal~\cite{guo2019simple} introduced a simple black-box attack (SimBA) which decides the direction of the perturbations based on the changes of output probability. Brendel \etal~\cite{brendel2017decision} first proposed a decision-based attack. Based on this method, Cheng \etal~\cite{cheng2018query} and Cheng \etal~\cite{chen2019hopskipjumpattack} improved the query efficiency, which is an important metric for black-box attacks. 
\vspace{-1em}
\paragraph{Adversarial Defenses}
Several defense methods for increasing the robustness of models have been proposed. Adversarial training~\cite{szegedy2013intriguing, madry2018towards, kurakin2016adversarialm,tramer2018ensemble} modifies the training schemes of the models, they directly train with the adversarial examples. Another method aims to modify the adversarial examples themselves such as random transformation ~\cite{kurakin2016adversarial, meng2017magnet, xie2018mitigating}. Buckman \etal~\cite{buckman2018thermometer} proposed a nonlinear transformation based on one-hot encoding to inputs of models. Gradient masking methods~\cite{tramer2018ensemble, dhillon2018stochastic} destroy the gradient information so that they fail the optimization-based attacks. However, these defense methods based on gradient masking have been showed unreliable~\cite{athalye2018obfuscated}, and models with defenses above are still unsafe against some attacks~\cite{carlini2017adversarial, he2017adversarial}. Besides, detecting adversarial examples raises the interest of researchers. Some of them detect the examples of whether they are adversarial or clean by an auxiliary network~\cite{gong2017adversarial, grosse2017statistical, metzen2017detecting}, while some find out adversarial examples through their statistical properties~\cite{bhagoji2017dimensionality, hendrycks2017early, feinman2017detecting, ma2018characterizing, papernot2018deep}.

\section{Method}
In this section, we describe the attack scenario in this study, then introduce the substitute attack and propose a data-free method to train the substitute model.

\subsection{Attack Scenario}
\paragraph{Label-only scenario} Suppose the attacked machine learning model is employed online and attackers can freely probe the output labels of the attacked model. The attackers are hard to obtain any data which is in the input space of the attacked model. We name the proposed DaST on the label-only scenario as DaST-L.

\paragraph{Probability-only scenario} The other settings of this scenario are the same as the label-only scenario, but attackers can access the output probability of the attacked model. We name the proposed DaST on the probability-only scenario as DaST-P.

\subsection{Adversarial Attack}
In this subsection, we introduce the definition of adversarial substitute attacks.

$\mathbf{X}$ denotes samples from the input space of the attacked model $T$. $\bar{y}$ and $y'$ refers to the real labels and the target labels of the samples $\mathbf{X}$, respectively. $T(y|\mathbf{X}, \theta)$ is the attacked model parameterized by $\theta$. For non-targeted attacks, the objective of the adversarial attack can be formulated as:

\begin{align}
\label{nontargeted}
\begin{split}
\underset{{\mathbf{\epsilon}}}{\text{min}} {\lVert {\epsilon} \rVert} \ \ \text{subject to} & \ \ \underset{{y_i}}{\text{argmax}} \  {T(y_i|\overline{\mathbf{X}} \! =  \! \mathbf{X}  \!  +  \! \mathbf{\epsilon}, \theta)} \ne \bar{y} \\
% \mathbf{s.t.}\quad
%& R = \|\mathbf{W} \|_0, \\
\text{and} &   \ \ \lVert \epsilon \rVert \le r.
%& & & R = \|\mathbf{W} \|_0\\
\end{split}
\end{align}
For targeted attacks, the objective is:

\begin{align}
\label{targeted}
\begin{split}
\underset{{\mathbf{\epsilon}}}{\text{min}} {\lVert {\epsilon} \rVert} \ \  \text{subject to} & \ \  \underset{{y_i}}{\text{argmax}} \  {T(y_i|\overline{\mathbf{X}} \! = \! \mathbf{X} \! + \! \mathbf{\epsilon}, \theta)} = y' \\
% \mathbf{s.t.}\quad
%& R = \|\mathbf{W} \|_0, \\
\text{and} &  \ \  \lVert \epsilon \rVert \le r,
%& & & R = \|\mathbf{W} \|_0\\
\end{split}
\end{align}
where the $\epsilon$ and $r$ are perturbation of the sample and upper bound of the perturbation, respectively. For attacking the machine learning system which is hard to detect, $r$ is set to a small value in attack methods. $\overline{\mathbf{X}} = \mathbf{X} + \mathbf{\epsilon}$ are the adversarial examples which can lead the attacked model $T$ to output a wrong label (non-targeted setting) and a specific wrong label (targeted setting). 

\begin{figure*}
	\begin{center}
		\includegraphics[width=0.95\textwidth]{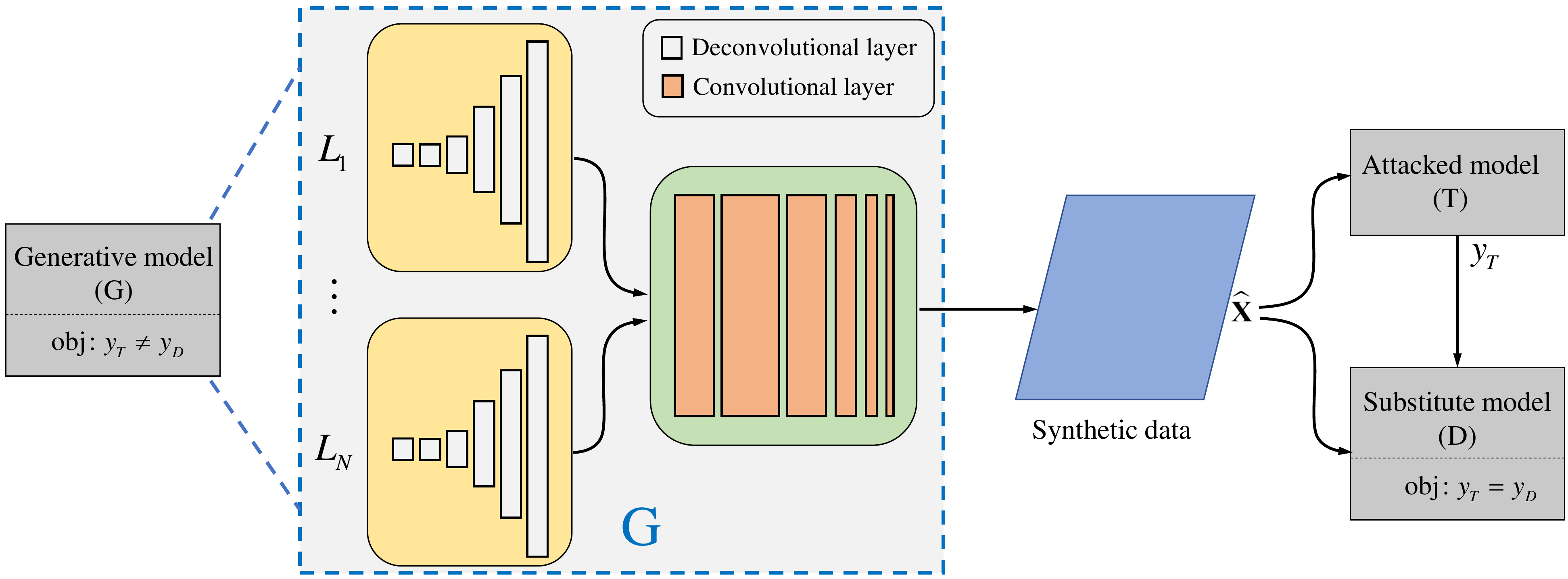}
	\end{center}
	\caption{The proposed adversarial data-free imitation. The architecture of G is shown in the blue dotted block. $N$ denotes the number of categories. In the training stage, the objective of $G$ is to generate samples $\widehat{\mathbf{X}} = G(\mathbf{X})$ and let $y_D(\widehat{\mathbf{X}}) \ne y_T(\widehat{\mathbf{X}}) $. The objective of $D$ is to guarantee $y_D(\widehat{\mathbf{X}}) = y_T(\widehat{\mathbf{X}}) $. In the testing stage, the substitute model $D$ is utilized to generate adversarial examples to attack $T$.}
	\label{fig:framework}
\end{figure*}

For white-box attacks, they can fully access the gradient information of $T$, then use it to generate adversarial examples to attack the $T$. For black-box substitute attacks, they train a model  $\widehat{T}$ to substitute the attacked model to generate adversarial examples and then transfer the examples to attack the $T$. The attack success rate of these black-box attacks heavily relies on the transferability of the adversarial examples. Therefore, the key point of developing an efficient substitute attack is to train a substitute model having properties that are as similar as possible to the attacked model. Current attack methods utilize the same training set of the attacked model or collect a lot of other images labeled by the attacked model to train the substitute model. In the next two subsections, we will introduce a method that can train a substitute model without any image. The whole process is shown in Figure \ref{fig:framework}. 

\subsection{Adversarial Generator-Classifier Training}
In this subsection, we introduce the basic adversarial training method and discuss its limitation.

For training the substitute model without any image, we use a generative model $G$ to produce training data for the substitute model $D$. The generator randomly samples the noise vector $\mathbf{z}$ from the input space and produces the data $\widehat{\mathbf{X}} = G({\mathbf{z}})$. Then, the generated data is used to probe the output $T(\widehat{\mathbf{X}})$ of the attacked model $T$. The substitute model is trained by the image-output pair $(\widehat{\mathbf{X}}, T(\widehat{\mathbf{X}}))$. As shown in Figure \ref{fig:framework}, the objective of $G$ is to create new samples to explore the difference between $T$ and $D$, and the role of $D$ is to imitate the output of $T$. It is a special two-player game, the attacked model involved in this game is a referee. To simplify the expression but without loss of generality, we utilize the binary classification as a case to analyze (the output probability can be considered as one scalar in binary classification, so does the output label). The value function of the game is presented as:
\begin{align}
\label{formular:game}
\begin{split}
\underset{G}{\text{max}} \ \underset{D}{\text{min}} \ \ &\mathcal{V}_{G, D} =  d(T(\widehat{\mathbf{X}}), D(\widehat{\mathbf{X}}))\\
\end{split}
\end{align}
where $d(T(\widehat{\mathbf{X}}), D(\widehat{\mathbf{X}}))$ is a metric to measure the output distance between $T$ and $D$. For label-only attack scenario, this measurement can be formulated as:
\begin{align}
\label{formular:label}
\begin{split}
d(T, D)  = \text{CE}(D(\widehat{\mathbf{X}}), T(\widehat{\mathbf{X}})), \\
\end{split}
\end{align}
where $D(\widehat{\mathbf{X}})$ and $ T(\widehat{\mathbf{X}})$ in this scenario denote the output labels of the substitute model and those of the attacked model, respectively. $\text{CE}(D(\widehat{\mathbf{X}}), T(\widehat{\mathbf{X}}))$ denotes the cross entropy loss, and the output labels of $T$ are utilized as the label of this loss. The function of cross entropy loss is to constrain the difference between the $T$ and $D$. For probability-only attack scenario, this measurement is formulated as:
\begin{align}
\label{formular:proba}
\begin{split}
d(T, D)  =  \| D(\widehat{\mathbf{X}}), T(\widehat{\mathbf{X}}) \|_F, \\
\end{split}
\end{align}
where $D(\widehat{\mathbf{X}})$ and $ T(\widehat{\mathbf{X}})$ in this scenario denote the output probabilities of the substitute model and those of the attacked model, respectively.

Hence the substitute model $D$ replicates the information of attacked model $T$ by this adversarial training. In the training, the loss function of $D$ is set to $\mathcal{L}_{D} = \mathcal{V}_{G, D}$. In order to maintain the stability of training, the loss function of $G$ is designed as $\mathcal{L}_{G} = e^{-d(T, D)}$. Therefore, the global optimal substitute network $D$ is obtained if and only if $ \ \forall \ \widehat{\mathbf{X}}, \ T(\widehat{\mathbf{X}})  =  D(\widehat{\mathbf{X}}) $. At this point, $\mathcal{L}_{D} = 0$ and $\mathcal{L}_{G} = e^{0} = 1$.

We suppose that $\forall \ \widehat{\mathbf{X}} = G({\mathbf{z}}), \widehat{\mathbf{X}} \in \mathbb{R} $, $\mathbb{R}$ is the input space of $T$. If the $D$ can achieve $D(\widehat{\mathbf{X}}) = T(\widehat{\mathbf{X}})$, the adversarial attacks carried out by our substitute model will have the same success rate as the white-box attack without the gradient information of $T$. Therefore, for a well-trained substitute network, adversarial examples generated by $D$ have strong transferability for $T$.

However, it is impossible to guarantee that $D(\widehat{\mathbf{X}}) = T(\widehat{\mathbf{X}})$ in a limited time. If we do not constrain the output of $G$, the synthetic training data for $T$ is likely only distributed in a small range of $\mathbb{R}$, thus this training cannot work. For addressing this problem, we design a label-controllable architecture for $G$, which can control the distribution of synthetic data and speed up the convergence of training.

\subsection{Label-controllable Data Generation}
In this subsection, we introduce the label-controllable architecture for the generative model $G$. 

To obtain equally distributed synthetic data to train the substitute model $D$, we consider developing a method that can control the distribution of $\widehat{\mathbf{X}}$. For training a replication of $T$, the synthetic data is used to probe the information of the attacked model. The label of samples, which is produced by the attacked model, should span all categories. Therefore, as shown in the blue dotted box of Figure \ref{fig:framework}, we design a generative network which contains $N$ upsampling deconvolutional components, $N$ is the number of categories. All upsampling components share a post-processing convolutional network. The model $G$ randomly samples the noise vector $\mathbf{z}$ from the input space and variable label value $n$. The $\mathbf{z}$ is then entered into the $n$-th upsampling deconvolutional network and the shared convolutional network to produce the data $\widehat{\mathbf{X}} = G({\mathbf{z}}, n)$. The additional label-control loss for generative model $G$ is formulated as:

\begin{align}
\label{formular:label-control}
\begin{split}
\mathcal{L}_C = \text{CE}(T(G({\mathbf{z}}, n)), n). \\
\end{split}
\end{align}

The above method generates data with random labels, which are produced by $T$. However, the back propagation of this label-control loss needs the gradient information of the attacked model $T$, it violates the rules of black-box attacks. We need to train a label-controllable generative model without the gradient information of $T$. For the imitation process, it can be approximated as the following objective function:

\begin{align}
\label{formular:imitation process}
\begin{split}
\underset{D}{\text{min}} \quad d(T(\widehat{\mathbf{X}}), D(\widehat{\mathbf{X}})). \\
\end{split}
\end{align}

In the training progresses, the outputs of $D$ will gradually approach the outputs of $T$ under the same inputs. Therefore, we use $D$ to replace the $T$ in Eq. (\ref{formular:label-control}), which is formulated as:
\begin{align}
\label{formular:label-controlnew}
\begin{split}
\mathcal{L}_C = \text{CE}(D(G({\mathbf{z}}, n)), n). \\
\end{split}
\end{align}
The training of substitute $D$ can avoid accessing the information of $T$. Then we update the loss of $G$ as:
\begin{align}
\label{formular:lossG}
\begin{split}
\mathcal{L}_G =e^{-d(T, D)} + \alpha \mathcal{L}_C, \\
\end{split}
\end{align}
where $\alpha$ controls the weight of label-control loss (we set it to 0.2 in our experiments).

In the training stage, as the imitation ability of $D$ increases, the diversity of synthetic samples which is labeled by the $T$ will enhance. Therefore, the $D$ can learn the information of the attacked model $T$, which can improve the transferability of adversarial examples generated by $D$. We name this method as data-free substitute training (DaST), which is shown in Algorithm 1.

\begin{table}[h]
	\centering
	\begin{tabular}{lll}
		\toprule
		$\bf Algorithm~1 $ Mini-batch stochastic gradient descent \\ training of the proposed method DaST.\\
		\midrule
		% 		\textbf{Input:}\\
		% 		%		$ \alpha, \beta: \text{the distance threshold of pixel-based matching and }$ \\ \text{feature-based matching, respectively.}  \\
		% 		$ \# \  \mathbf{F}, \mathbf{F}_i: \text{are the characteristics extracted by the current }$ \\ $\text{input image and the i-th previous input image. }$\\
		% 		$ \# \  \mathbf{X}^*_{n+1,i}: \text{the stacked input of } \mathbf{X}_{n+1} \text{ and} \mathbf{X}_{i}. $ \\
		% 		\textbf{Output:}\\
		$\# \  acc \text{ denotes the accuracy of D. } att \text{ denotes the attack} $ \\
		$ \text{success rate for the attacks generated by D.} $ \\
		$1: \textbf{While} \ \text{iteration} < \delta \textbf{ or }  acc, att \text{ do not increace}  $ \\
		$2: \qquad \text{Generate } m \text{ examples} \{ \widehat{\mathbf{X}}^{(1)}, \ldots , \widehat{\mathbf{X}}^{(m)}  \} \text{by } G. $ \\
		$3: \qquad \text{Update the substitute model}: $\\
		$4: \qquad \qquad \mathcal{L}_{D} = d(T(\widehat{\mathbf{X}}), D(\widehat{\mathbf{X}})).  $ \\
		% 		$5: \textbf{else} $  \\
		$5: \qquad \text{Update the generative model}: $     \\
		$6: \qquad \qquad \mathcal{L}_G =e^{-d(T, D)} + \alpha \mathcal{L}_C.  $   \\
		$7: \textbf{end for}  $ \\
		%		$ \#  \ \   \text{We set } \alpha \text{ to } 2 \times 10^{-5} \text{ in our experiments}$. \\
		\bottomrule	
	\end{tabular}
	\label{adt}
	\vspace{-0.5em}
\end{table}

Like the current substitute attack methods, the substitute model trained by our method is utilized to generate adversarial examples to attack $T$. 

\section{Experiments}
\subsection{Experiment Setting}
In this subsection, we introduce our experiment settings, including the datasets, model architectures, attack methods, and evaluation criteria. 
\vspace{-1em}
\paragraph{Dataset:} we evaluate our proposed method on MNIST \cite{lecun1998gradient} and CIFAR-10 \cite{krizhevsky2009learning}. The test sets of these two datasets have 10k images, respectively.
\vspace{-1em}
\paragraph{Scenario:} we evaluate our method in both label-only attack and probability-only scenario. The DaST-L and DaST-P denote the DaST in the label-only scenario and DaST in the probability-only scenario, respectively. Attackers in the scenarios of this study can freely access the output of the attacked model. Therefore, we obtain the substitute model trained by DaST when the algorithm convergence.
\vspace{-1em}
\paragraph{Model architecture and attack method:} the substitute network has no prior knowledge of the attacked model, which means it does not load any pre-trained model in experiments. For the experiments on MNIST, we design 3 different network architectures including a small network (3 convolutional layers), a medium network (4 convolutional layers) and a large network (5 convolutional layers) for evaluating the performance of our DaST with models having different capacity. We utilize the pre-trained medium network and VGG-16 \cite{Karen2015vgg} as the attacked model on MNIST and CIFAR-10, respectively. In addition, we use different architectures for the substitute model and attacked model to evaluate the impact of model structure on our method in CIFAR-10 experiments. In order to compare the substitute model produced by DaST with the pre-trained models, we utilize 4 attack methods to generate adversarial examples, which include FGSM~\cite{goodfellow6572explaining}, BIM~\cite{kurakin2016adversarial}, projected gradient descent (PGD)~\cite{madry2018towards}, C\&W~\cite{carlini2017towards}. For testing, we use AdverTorch library \cite{ding2018advertorch} to generate adversarial examples. For evaluating performances of the proposed method in real-world tasks, we apply our attack to the online MNIST model of Microsoft Azure. The training tricks and machine learning methods utilized by this online model cannot be accessed.  
\vspace{-1em}
\paragraph{Evaluation criteria:} for evaluating the performance of our DaST, we set the attack success rates of adversarial examples generated by other pre-trained networks as the baseline. The goals of non-targeted attacks and targeted attacks are to lead the attacked model to output wrong labels and specific wrong labels, respectively. In the non-targeted attack scenario, we only generate adversarial examples on the images classified correctly by the attacked model. In targeted attacks, we only generate adversarial examples on the images which are not classified to the specific wrong labels. The success rates of adversarial attack are calculated by $ n / m$, where $n$ and $m$ are the number of adversarial examples which can fool the attacked model and the total number of adversarial examples, respectively.

\subsection{Experiments on MNIST}

In this subsection, we employ the proposed DaST to train a substitute model for adversarial attacks on the MNIST dataset and evaluate the performance in terms of attack success rate in label-only and probability-only scenarios.

\begin{table} [t]
	%	\vspace{-1em}
	\caption{Performance of the proposed DaST on MNIST. ``Pre-trained'', ``DaST-L'' and ``DaST-P'': the attack success rate ($\%$) of adversarial examples generated by the pre-trained large network and DaST-L and DaST-P, respectively. ( ) denotes the average $L_F$ perturbation distance per image.}
	\label{minist_attack}
	%	\small
	\centering
	\vspace{+0.5em}
	%	\vspace{+0.5em}
	\begin{tabular}{lccc}
		\toprule
		\multirow{2}{*}{Attack} 
		&\multicolumn{3}{c}{\textbf{Non-targeted}  }  \\
		
		\cline{2-4} 
		\specialrule{0em}{1pt}{1pt}
		&Pre-trained &DaST-P  &DaST-L\\
		\midrule
		% non-target: wb, bb, im                                  % target: wb, bb, im
		FGSM \cite{goodfellow6572explaining}  &59.72 (5.40)  &\textbf{69.76} (5.41)   & 35.74 (5.40)\\
		% \midrule
		BIM \cite{kurakin2016adversarial}  &85.70 (4.80)  &\textbf{96.36} (4.81)   & 64.61 (4.82)\\
		% \midrule
		PGD \cite{madry2018towards}  &37.93 (3.98)  &\textbf{53.99} (3.99)   & 23.22 (3.98)\\
		% \midrule
		C\&W  \cite{carlini2017towards}  &23.34 (2.91)  &\textbf{27.35} (2.74)   & 18.16 (2.75)\\
		
		\toprule
		\multirow{2}{*}{Attack} 
		&\multicolumn{3}{c}{\textbf{Targeted} } \\
		
		\cline{2-4} 
		\specialrule{0em}{1pt}{1pt}
		&Pre-trained &DaST-P  &DaST-L\\
		\midrule
		% non-target: wb, bb, im                                  % target: wb, bb, im
		FGSM \cite{goodfellow6572explaining} &12.10 (5.46)  &\textbf{20.45} (4.49) & 13.10 (5.46)\\
		% \midrule
		BIM \cite{kurakin2016adversarial}  &37.83 (4.90)  &\textbf{57.22} (4.87) & 29.18 (4.87)\\
		% \midrule
		PGD \cite{madry2018towards}  &28.95 (4.60)  &\textbf{47.57} (4.63) & 19.25 (4.63)\\
		% \midrule
		C\&W   \cite{carlini2017towards}  &10.32 (2.57)  &\textbf{23.80} (2.99) & 12.31 (2.98) \\
		
		\bottomrule
	\end{tabular}
	\vspace{-0.5em}
\end{table}

First, we conduct experiments to evaluate the performance in probability-only and label-only attack scenarios. We use the medium network as the attacked model on MNIST and the large network as the substitute model of DaST. We train a pre-trained large network on the same train set of the attacked model. We utilize the attack success rate of adversarial examples generated by the pre-trained model as the baseline. The performances of our DaST are shown in Table \ref{minist_attack}.The substitute model trained by DaST-P and DaST-L achieve 97.82\% and 83.95\% of accuracy on the test set, respectively. The attack success rates of the substitute model produced by our DaST are higher than those of the pre-trained model on non-targeted (10.04\%, 10.66\%, 16.06\%, and 4.01\% higher on FGSM, BIM, PGD, and C\&W, respectively) and targeted attacks (11.83\%, 19.39, 18.62, 13.48\% higher on FGSM, BIM, PGD, and C\&W, respectively). It shows that the substitute model generated by DaST-P outperform the models trained by the same train set (60000 images) with the attacked model. Even he substitute models trained by DaST-L perform better than baseline models on FGSM and C\&W attacks (targeted).

\begin{table} [t]
	%	\vspace{-1em}
	\caption{Performances of the proposed DaST with three different substitute architectures on MNIST. ``Small'', ``Medium'' ``Large'': the attack success rates ($\%$) of adversarial examples generated by DaST with small, medium and large substitute networks, respectively. ( ) denotes the average $L_F$ perturbation distance per image.}
	\label{minist_differen_attack}
	%	\small
	\centering
	\vspace{+0.5em}
	%	\vspace{+0.5em}
	\begin{tabular}{lccc}
		\toprule
		\multirow{2}{*}{Attack} 
		&\multicolumn{3}{c}{\textbf{Non-targeted}  } \\
		
		\cline{2-4} 
		\specialrule{0em}{1pt}{1pt}
		&Small &Medium &Large \\
		\midrule
		% non-target: wb, bb, im                                  % target: wb, bb, im
		FGSM \cite{goodfellow6572explaining}  &62.61 (4.38)  &56.21 (4.45)  &\textbf{69.76} (5.41)   \\
		% \midrule
		BIM  \cite{kurakin2016adversarial}  &94.86 (4.85)  &92.47 (4.84)  &\textbf{96.36} (4.81) \\
		% \midrule
		PGD \cite{madry2018towards}  &45.31 (3.99)  &43.62 (3.99)  &\textbf{53.99} (3.99)  \\
		% \midrule
		C\&W   \cite{carlini2017towards}  &\textbf{30.61} (2.89)  &24.34 (2.75)  &23.80 (2.99)  \\
		\bottomrule
		\specialrule{0em}{1pt}{1pt}
		\multirow{2}{*}{Attack} 
		&\multicolumn{3}{c}{\textbf{Targeted}}  \\
		\cline{2-4} 
		\specialrule{0em}{1pt}{1pt}
		&Small &Medium &Large \\
		\midrule
		% non-target: wb, bb, im                                  % target: wb, bb, im
		FGSM \cite{goodfellow6572explaining}   &19.92 (4.43))  &20.45 (4.49)  &\textbf{23.93} (5.45)\\
		% \midrule
		BIM \cite{kurakin2016adversarial}    &56.73 (4.89)  &53.50 (4.84)  &\textbf{57.22} (4.87)\\
		% \midrule
		PGD  \cite{madry2018towards}  &39.42 (4.64)   &40.76 (4.60)  &\textbf{47.57} (4.63)\\
		% \midrule
		C\&W    \cite{carlini2017towards}  &\textbf{24.86} (3.09)  &16.25 (3.13)  &23.80 (2.99)\\
		
		\bottomrule
	\end{tabular}
	%	\vspace{-1em}
\end{table}

Then we evaluate the performances of our DaST with different substitute architectures in the probability-only scenario. We also use the medium network as the attacked model on MNIST and apply our DaST using three different substitute architectures, which include the large, medium and small networks. The attack success rates of these three substitute architectures are shown in Table \ref{minist_differen_attack}. The large substitute model achieves the best results on FGSM, BIM, PGD attacks compared with other models. The small substitute model obtains the best results on C\&W attacks compared with other models. It shows that both architectures for the substitute model obtain good results on adversarial attacks. In general, the substitute models with more complex structure can obtain better performance for adversarial attacks.

\subsection{Experiments on CIFAR-10}
In this subsection, we employ the proposed DaST to train a substitute model for adversarial attacks on the CIFAR-10 dataset, and evaluate the performance in terms of attack success rate in label-only and probability-only scenarios.

We conduct experiments to evaluate the performance in probability-only and label-only attack scenarios and use the VGG-16 network as the attacked model. We train a pre-trained ResNet-50 network on the same train set of the attacked model. The performances of our DaST are shown in Table \ref{cifar_attack}. The substitute model trained by DaST-P and DaST-L achieve 25.15\% and 20.35\% of accuracy on the test set, respectively. Our DaST also achieves competitive performance with the pre-trained model. In most cases of the probability-only scenario (FGSM, BIM, C\&W for non-targeted attack, BIM, PGD, C\&W for targeted attacks), the substitute models generated by DaST-P outperform baseline models. The substitute models trained by DaST-L perform better than baseline models on C\&W attacks (non-targeted).

\begin{table} [t]
	%	\vspace{-1em}
	\caption{Performance of the proposed DaST on CIFAR-10. ``Pre-trained'', ``DaST-P'' ``DaST-L'': the attack success rates ($\%$) of adversarial examples generated by the pre-trained large network, DaST-P and DaST-L, respectively. ( ) denotes the average $L_F$ perturbation distance per image.}
	\label{cifar_attack}
	%	\small
	\centering
	%	\vspace{+0.5em}
	\vspace{+0.5em}
	\begin{tabular}{lccc}
		\toprule
		\multirow{2}{*}{Attack} 
		&\multicolumn{3}{c}{\textbf{Non-targeted}  } \\
		
		\cline{2-4} 
		\specialrule{0em}{1pt}{1pt}
		&Pre-trained &DaST-P  &DaST-L\\
		\midrule
		% non-target: wb, bb, im                                  % target: wb, bb, im
		FGSM \cite{goodfellow6572explaining}  &39.10 (1.54)  &\textbf{39.63} (1.54)   & 22.65 (1.54)\\
		% \midrule
		BIM  \cite{kurakin2016adversarial}  &59.18 (1.01)  &\textbf{59.71} (1.18)   & 28.42 (1.19)\\
		% \midrule
		PGD  \cite{madry2018towards} &\textbf{35.40} (1.02)  &29.10 (1.10)   & 17.80 (1.10)\\
		% \midrule
		C\&W  \cite{carlini2017towards}   &9.76 (0.77)  &\textbf{13.52} (0.74)   & 10.34 (0.74)\\
		
		\toprule
		\multirow{2}{*}{Attack} 
		&\multicolumn{3}{c}{\textbf{Targeted}} \\
		
		\cline{2-4} 
		\specialrule{0em}{1pt}{1pt}
		&Pre-trained &DaST-P  &DaST-L\\
		\midrule
		% non-target: wb, bb, im                                  % target: wb, bb, im
		FGSM \cite{goodfellow6572explaining}  &\textbf{9.62} (1.54)  &6.69 (1.54)   & 7.32 (1.54)\\
		% \midrule
		BIM \cite{kurakin2016adversarial}   &17.43 (1.00)  &\textbf{20.22} (1.18) & 15.26 (1.16)\\
		% \midrule
		PGD \cite{madry2018towards}  &10.46 (1.05)  &\textbf{14.09} (1.12) & 8.32 (1.10)\\
		% \midrule
		C\&W  \cite{carlini2017towards}   & 23.15 (2.05)  &\textbf{26.53} (1.98) & 19.78 (2.04)\\
		
		\bottomrule
	\end{tabular}
	%	\vspace{-1em}
\end{table}

\begin{figure}
	\vspace{+0.5em}
	\begin{center}
		\includegraphics[width=0.46\textwidth]{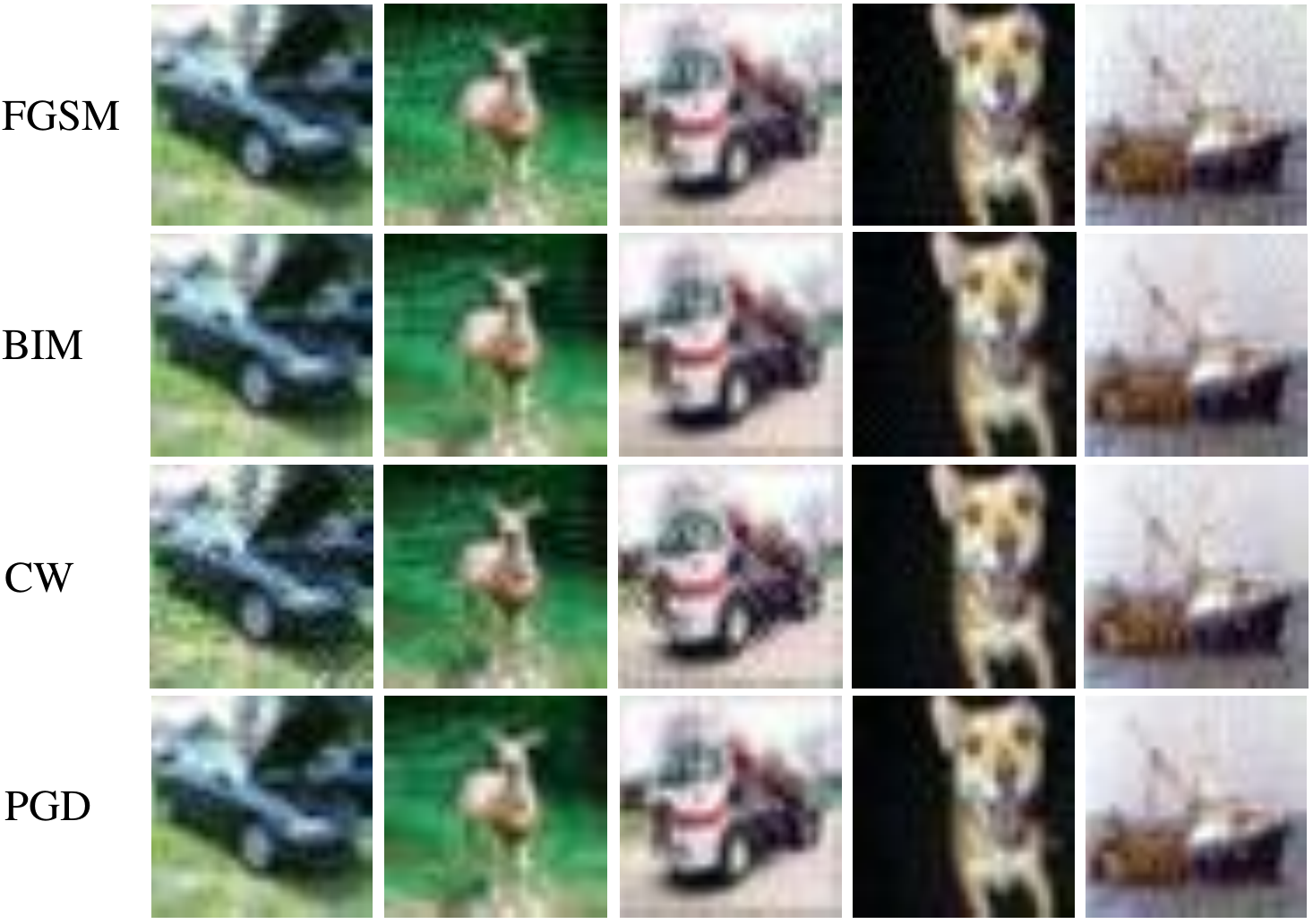}
	\end{center}
	\caption{Visualization of the adversarial examples generated by DaST-L on CIFAR-10. We generate 5 samples for each attack.}
	\label{fig:l}
\end{figure}

\begin{figure}
	\vspace{+0.5em}
	\begin{center}
		\includegraphics[width=0.46\textwidth]{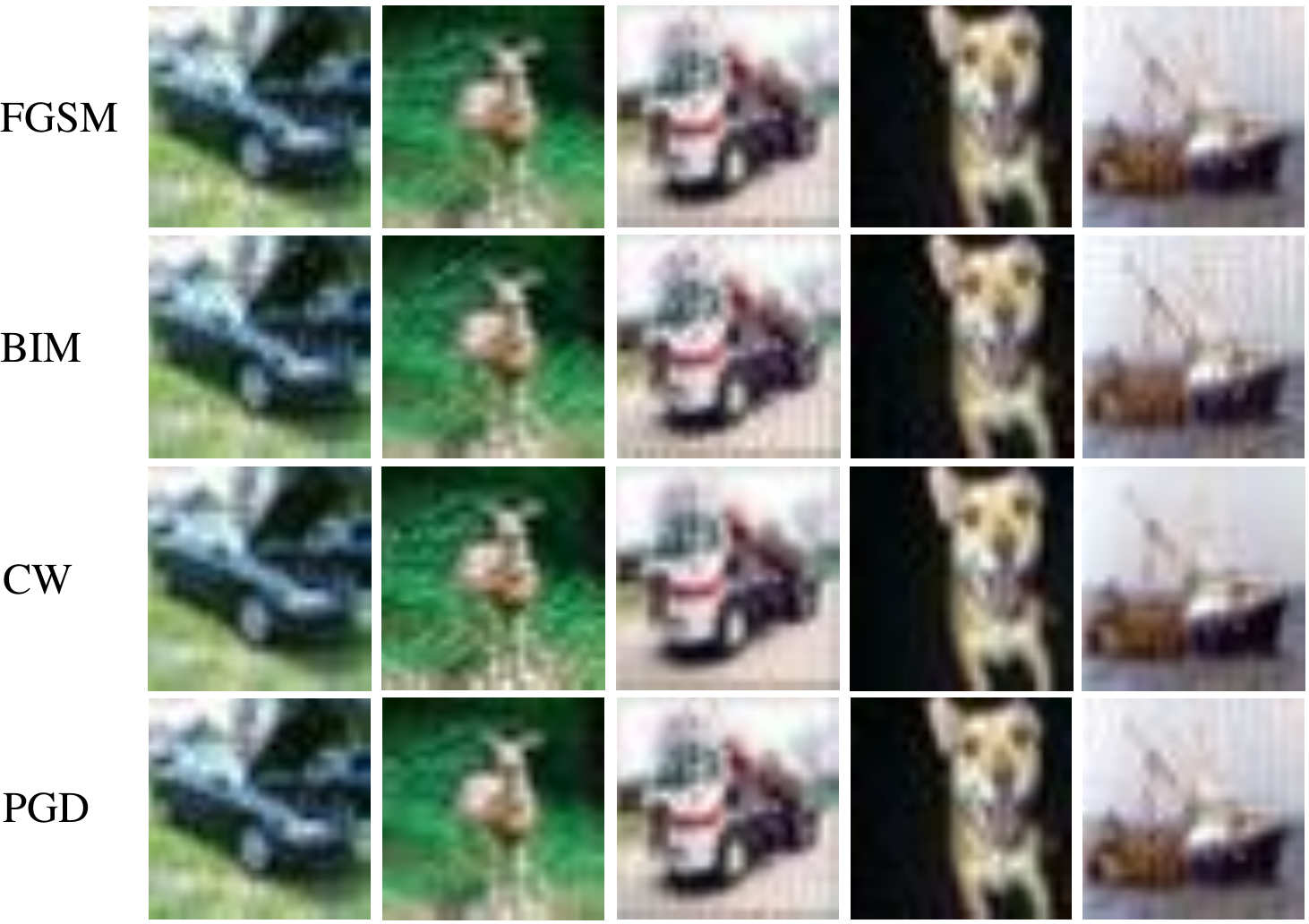}
	\end{center}
	\caption{Visualization of the adversarial examples generated by DaST-P on CIFAR-10. We generate 5 samples for each attack.}
	\label{fig:p}
\end{figure}

We also evaluate the performances of our DaST with different substitute architectures in the probability-only scenario. The VGG-16 network is used as the attacked model. We apply our DaST using 3 different substitute architectures, which include the VGG-13, ResNet-18, and ResNet-50. The attack success rates of these three substitute architectures are shown in Table \ref{cifar_differen_attack}. It demonstrates that both architectures for the substitute model obtain good results on adversarial attacks. In most cases (BIM, PGD, C\&W for non-targeted attack, FGSM, BIM, PGD, C\&W for targeted attacks), the VGG-13 outperforms other models in terms of the adversarial attack. The ResNet-50 obtains the best results on FGSM attacks (targeted). Different from experiments on MNIST, the simple model achieves the best results on CIFAR-10. We visualize the adversarial examples generated by DaST-P and DaST-L in Figure \ref{fig:l} and \ref{fig:p}, respectively. The attack perturbations for these two scenarios are small. 

\begin{table} [t]
	%	\vspace{-1em}
	\caption{Performances of the proposed DaST with three different substitute architectures on CIFAR-10. ``VGG-13'', ``ResNet-18'' ``ResNet-50'': the attack success rates (the high is better) of adversarial examples generated by DaST with VGG-13, ResNet-18 and ResNet-50 substitute models, respectively. The numbers in ( ) denote the average $L_F$ perturbation distance per image.}
	\vspace{+0.5em}
	\label{cifar_differen_attack}
	%	\small
	\centering
	%	\vspace{+0.5em}
	\begin{tabular}{lccc}
		\toprule
		\multirow{2}{*}{Attack} 
		&\multicolumn{3}{c}{\textbf{Non-targeted} ($\%$) } \\
		
		\cline{2-4} 
		\specialrule{0em}{1pt}{1pt}
		&VGG-13 &ResNet-18 &ResNet-50 \\
		\midrule
		% non-target: wb, bb, im                                  % target: wb, bb, im
		FGSM \cite{goodfellow6572explaining}  &6.87 (1.54)  &17.97 (1.54)  &\textbf{39.63} (1.54)   \\
		% \midrule
		BIM  \cite{kurakin2016adversarial}  &\textbf{93.13} (1.18)  &31.70 (1.54)  &59.71 (1.18) \\
		% \midrule
		PGD \cite{madry2018towards}  &\textbf{56.14} (1.08)  &10.04 (1.11)  &29.10 (1.10)  \\
		% \midrule
		C\&W   \cite{carlini2017towards}  &\textbf{56.80} (1.64)  & 11.54 (1.64)  &13.52 (0.74)  \\
		\bottomrule
		\specialrule{0em}{1pt}{1pt}
		\multirow{2}{*}{Attack} 
		&\multicolumn{3}{c}{\textbf{Targeted} ($\%$)}  \\
		\cline{2-4} 
		\specialrule{0em}{1pt}{1pt}
		&VGG-13 &ResNet-18 &ResNet-50 \\
		\midrule
		% non-target: wb, bb, im                                  % target: wb, bb, im
		FGSM \cite{goodfellow6572explaining}   &\textbf{18.27} (1.54)  &2.07 (1.54)  &6.69 (1.54)\\
		% \midrule
		BIM  \cite{kurakin2016adversarial}   &\textbf{62.23} (1.24)  &8.00 (1.52)  &20.22 (1.18)\\
		% \midrule
		PGD  \cite{madry2018towards}  &\textbf{41.48} (1.17)   &3.72 (1.26)  &14.09 (1.12)\\
		% \midrule
		C\&W   \cite{carlini2017towards}   &\textbf{33.65} (2.42)  & 7.31 (1.46)  &26.53 (1.98)\\
		
		\bottomrule
	\end{tabular}
	%	\vspace{-1em}
\end{table}

\begin{table} 
	%	\vspace{-1em}
	\caption{Performance of the proposed DaST for attacking Microsoft Azure example model. ``Pre-trained'', ``DaST-P'' ``DaST-L'': the attack success rate (the high is better) of adversarial examples generated by the pre-trained large network, DaST in probability-only scenario and DaST in label-only scenario, respectively. The numbers in ( ) denote the average $L_F$ perturbation distance per image. Because it hard to generate adversarial examples for all methods on C\&W \cite{carlini2017towards}, we omit this attack method.}
	\vspace{+0.5em}
	\label{azure_attack}
	%		\small
	\centering
	\vspace{+0.3em}
	\begin{tabular}{lccc}
		\toprule
		\multirow{2}{*}{Attack} 
		&\multicolumn{3}{c}{\textbf{Non-targeted} ($\%$) }\\
		
		\cline{2-4} 
		\specialrule{0em}{1pt}{1pt}
		&Pre-trained &DaST-P  &DaST-L \\
		\midrule
		% non-target: wb, bb, im                                  % target: wb, bb, im
		FGSM \cite{goodfellow6572explaining} & 77.96 (5.41)  &96.83 (5.25)  &\textbf{98.21} (5.36)  \\
		% \midrule
		BIM  \cite{kurakin2016adversarial} & 66.25 (4.81)  &96.42 (4.79)  &\textbf{98.35} (4.72)  \\
		% \midrule
		PGD \cite{madry2018towards}  & 59.23 (3.99)  &90.63 (3.88)  &\textbf{96.97} (3.96) \\
		
		\toprule
		\multirow{2}{*}{Attack} 
		&\multicolumn{3}{c}{\textbf{Targeted} ($\%$)} \\
		
		\cline{2-4} 
		\specialrule{0em}{1pt}{1pt}
		&Pre-trained &DaST-P  &DaST-L\\
		\midrule
		% non-target: wb, bb, im                                  % target: wb, bb, im
		FGSM \cite{goodfellow6572explaining}  & 13.52 (5.46) &32.00 (5.21)  &\textbf{43.99} (5.37)\\
		% \midrule
		BIM \cite{kurakin2016adversarial}  & 19.31 (4.88)  &50.21 (4.90)  &\textbf{71.15} (4.56)\\
		% \midrule
		PGD \cite{madry2018towards}  & 19.31 (4.60)  &45.66 (4.46)  &\textbf{65.91} (4.32)\\
		% \midrule
		%		C\&W    &23.34 (2.91)  &\textbf{27.35} (2.74)  &39.10 (1.54)  &\textbf{39.63} (1.54)\\
		
		\bottomrule
	\end{tabular}
	%	\vspace{-1em}
\end{table}

\subsection{Experiments on Microsoft Azure}
In this subsection, we conduct experiments for attacking the online model on Microsoft Azure in two scenarios.

We use the example MNIST model of the machine learning tutorial on Azure as the attacked model and employ it as a web service. We do not know the machine learning method and architecture of this model. The only information we can obtain is the outputs of this model. We apply the probability-based DaST and label-based DaST attacks to this model to evaluate the performance of the proposed method in real-world applications. The substitute model in this experiment has 5 convolutional layers. The substitute model trained by DaST-P and DaST-L achieve 79.35\% and 90.75\% of accuracy on the MNIST test set, respectively. The performance on adversarial attacks of the proposed method is shown in Table \ref{azure_attack}. 

\begin{figure} [t]
	\begin{center}
		\includegraphics[width=0.35\textwidth]{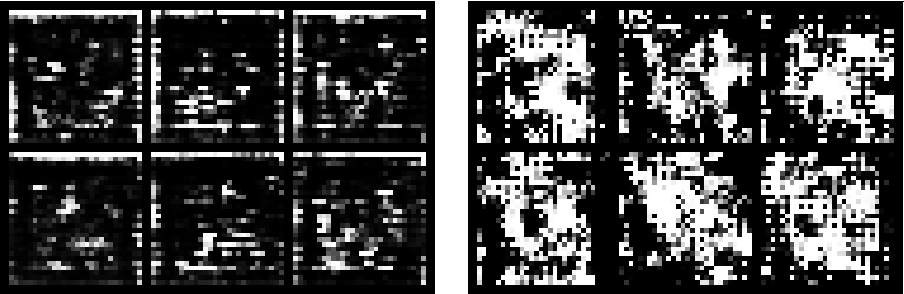}
	\end{center}
	\vspace{-0.5em}
	\caption{Visualization of the synthetic samples generated by the generator in the training of DaST. Left: samples generated by the DaST-L. Right: samples generated by the DaST-P.}
	\label{fig:visual}
\end{figure}

\begin{figure}
	\begin{center}
		\includegraphics[width=0.43\textwidth]{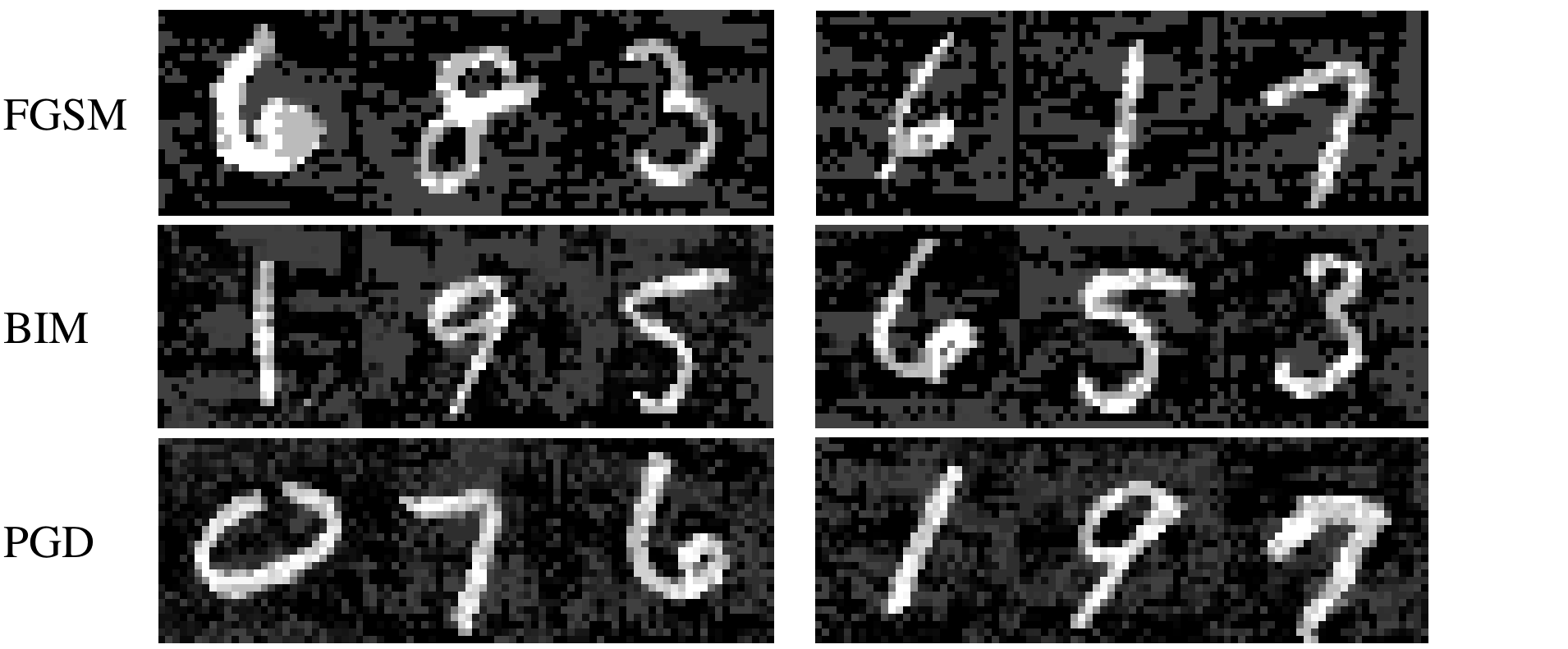}
	\end{center}
	\vspace{-0.5em}
	\caption{Visualization of the adversarial examples generated by DaST for attacking the Azure model. Left: examples generated by DaST-P. Right: examples generated by DaST-L.}
	%	 \vspace{-0.5em}
	\label{fig:exmaple}
\end{figure}

\begin{figure}
	\vspace{-1.2em}
	\begin{center}
		\includegraphics[width=0.4\textwidth]{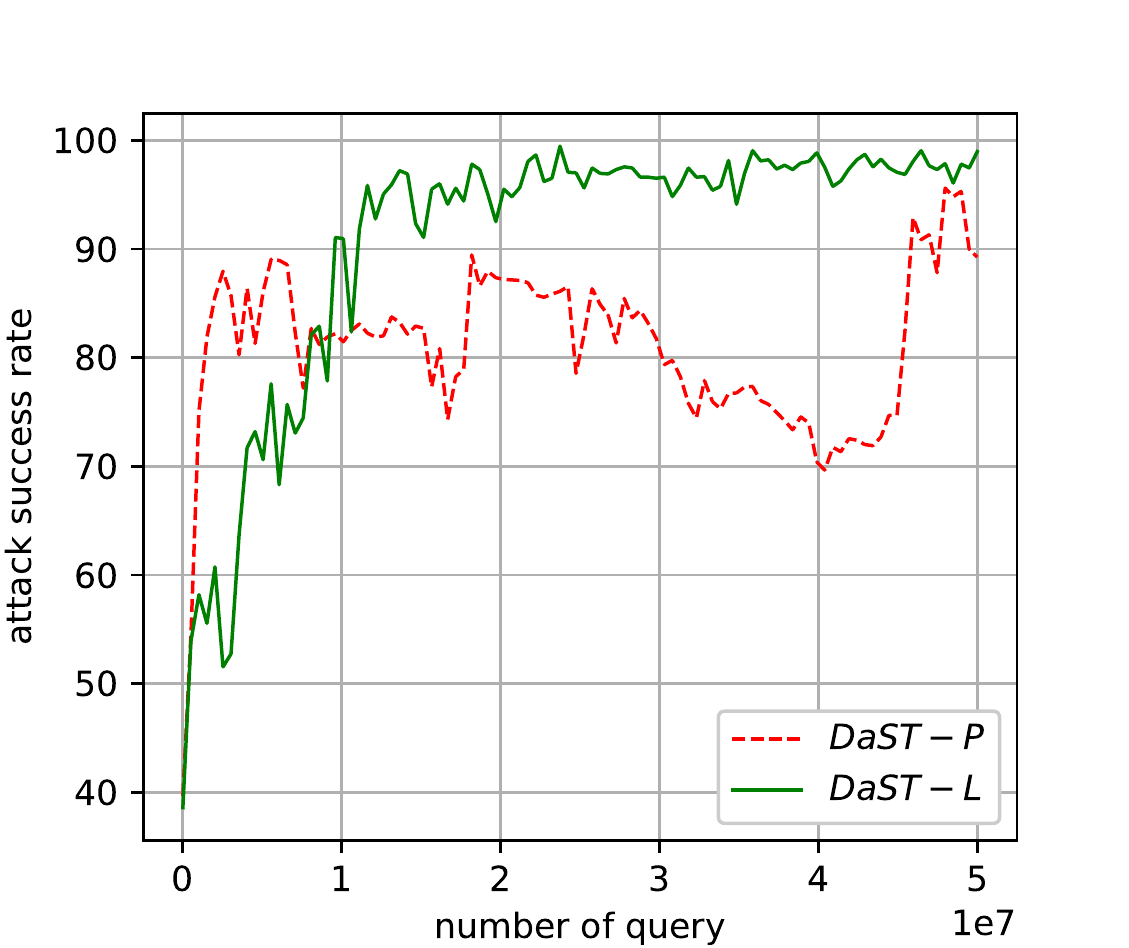}
	\end{center}
	\vspace{-0.5em}
	\caption{Attack success rate of BIM attacks generated by DaST in training stage of Azure experiments.} 
%	Note that the parameters of this BIM attack is different from the those in Table \ref{azure_attack}.}
%	\vspace{-0.5em}
	\label{fig:curve}
\end{figure}

The performance of DaST-L is better than its of DaST-P on this online model. Because the attacked Azure model is too simple, the accuracy on MNIST is only 91.93\%. Figure \ref{fig:curve} shows the training of DaST-P, which can access more information of attacked model than DaST-L, suffers over-fitting. DaST-L substitutes achieve a very high attack success rate on FGSM (98.21\%), BIM (98.35\%), PGD (96.97\%) attacks. Moreover, our DaST method achieves a high attack success rate even on targeted attacks. Compared with the models trained by the MNIST train set, substitute models trained by DaST perform much better on label-only (20.25\%, 32.10\%, 37.74\% higher on non-targeted FGSM, BIM, PGD attacks, respectively. 30.47\%, 51.84\%, 46.60\% higher on targeted FGSM, BIM, PGD attacks, respectively) and probability-only scenarios. It presents that our approach is better at attacking actual online models, even the proposed method does not need any real data. Because DaST does not need any query in the evaluation stage but needs queries in the training stage, our DaST requires different information than score-based attacks and decision-based attacks (they need queries in the evaluation stage). We show the number of queries for score-based and decision-based attacks, which have similar perturbation distance with DaST in non-targeted attacks. The results are shown in Table \ref{azure_attack_compare}. Our DaST is trained by 20,000,000 queries for the attacked model in the training stage. Compared with decision-based and score-based attacks, the input each time the DaST accesses the attacked model is different in the training stage (current query-based attacks need to use one original data to access the attacked model numerous times to generate each attack). So the queries of DaST are harder to be tracked than other attacks. 
\begin{table} 
	%	\vspace{-1em}
	\caption{Comparison of DaST and other attacks. "ASR": attack success rate. "Query": the number of queries in the evaluation stage. "Boundary": Decision-Based Attacks \cite{brendel2017decision}. "GLS": a score-based black-box attack based on greedy local search \cite{narodytska2016simple}. "-" denotes our DaST does not need query in the evaluation stage. The DaST in this experiment generate attacks with BIM.}
%	\vspace{+0.5em}
	\label{azure_attack_compare}
	%		\small
	\centering
	\vspace{+0.3em}
	\begin{tabular}{lccc}
		\toprule
		Attack & ASR & Distance & Query \\
		\specialrule{0em}{1pt}{1pt}
		\midrule
		% non-target: wb, bb, im                                  % target: wb, bb, im
		DaST-P & 96.83\%  &4.79  & -  \\
		% \midrule
		GLS  \cite{narodytska2016simple} & 40.51\%  &4.27  & 297.07  \\
		% \midrule
		DaST-L  & 98.35\%  &4.72  & - \\
		Boundary \cite{brendel2017decision}  & 100\%  &4.69  & 670.54 \\
		
		\bottomrule
	\end{tabular}
	%	\vspace{-1em}
\end{table}
\vspace{-1em}
\paragraph{Visualization:}we visualize the synthetic samples generated by the generative model in DaST on Azure experiments, which is shown in Figure \ref{fig:visual}. We also visualize the adversarial examples generated by DaST-P and DaST-L in Figure \ref{fig:exmaple}. The attack perturbations of DaST are small. 
\vspace{-1em}
\paragraph{Training convergence:} We show the curve of attack success rate of BIM attacks generated by DaST in the training stage of Azure experiments, which is shown in Figure \ref{fig:curve}. The attack success rates for DaST-L and DaST-P converge after 20,000,000 and 2,000,000 queries, respectively. 

%\begin{figure}
%	\begin{center}
%		\includegraphics[width=0.44\textwidth]{images/examples_l.pdf}
%	\end{center}
%	\caption{Visualization of the adversarial examples generated by DaST-P for attacking the Azure model. We generate 4 samples for each attack method. We only show the adversarial examples which can fool the model successfully.}
%	\label{fig:exmaples_l}
%\end{figure}  

\section{Conclusion}
We have presented a data-free method DaST to train substitute models for adversarial attacks. DaST reduces the prerequisites of adversarial substitute attacks by utilizing GANs to generate synthetic samples. This is the first method that can train substitute models without the requirement of any real data. The experiments showed the effectiveness of our method. It presented that machine learning systems have significant risks, attackers can train substitute models even when the real input data is hard to collect. 

The proposed DaST cannot generate adversarial examples alone, it should be used with other gradient-based attack methods. In future work, we will design a new substitute training method, which can generate attacks directly. Furthermore, we will explore the defense for DaST.

\section{Acknowledgment}
This research is supported by National Natural Science Foundation of China (NSFC, No. 61602091, No. 61571102, No.61872067) and Sichuan Science and Technology Program (No. 2019YFH0008, No.2018JY0035, No.2019YFH0016).

{\small
\bibliographystyle{ieee_fullname}
\bibliography{egbib}
}

\end{document}